\documentclass[amsmath,amssymb,lengthcheck,aps,prl] {revtex4}
\usepackage{graphics}
\usepackage{graphicx}

\begin{document}

\title{Collective excitations of two-dimensional Bose-Einstein Condensate in liquid phase with spin-orbit coupling}

\author{Saswata Sahu}
 \author{ Dwipesh Majumder }
\affiliation{Department of Physics, Indian Institute of Engineering Science and Technology, Shibpur, W B, India}
%\date{\today}

\begin{abstract}
We have studied the collective excitation of Bose-Einstein condensate of short-range weak interacting atoms with spin-orbit coupling (SOC) in two dimension in the liquid phase. In our study, we have included Rashba, Dresselhaus SOC, as well as Raman SOC. The study of Bogoliubov excitation shows that in small interaction, only phonon modes are present, and for higher interactions, roton modes start to appear.  Energy spectra contain two roton modes in the system relatively strong interacting atomic system.

\end{abstract}

\maketitle 

%droplets
Liquid formation of the dilute ultracold atomic system \cite{Petrov2015,Trarruell2018}  is one of the most exciting topics in Bose-Einstein condensation (BEC).
The droplets have been observed in the isotropic short-range interacting system of two species of cold atoms \cite{Trarruell2018,drop_exp2, Trarruell2018PRL} as well as in the anisotropic long-range dipolar interacting system of $^{164}$Dy or $^{166}$Er atoms \cite{dipolar_droplets}. 
In the mixture of two components Bose atoms, the spherical droplet has been observed under the competition between the effective short range attractive interaction and the repulsive interaction due to the quantum fluctuation\cite{LHY}.
The two component BEC may be the mixture of atoms of two different elements (different atomic mass)\cite{PRL'89, PRL'100} or maybe the mixture of atoms with two different internal degrees of freedom of a given element\cite{spin_BEC, PRL'101, 2Comp_BEC}. 
 In the dipolar system, the cigar-shaped droplet has been observed in the balanced attractive interaction due to the asymmetric dipolar interaction and repulsive interaction due to the quantum fluctuations.  Three-body collisions limit the lifetime of the droplets. In the lower dimensions, it is expected that this lifetime can be extended because of reduced phase-space available to colliding atoms. That's why people have an interest in droplets in the lower dimension.
There are already theoretical proposals of liquid states of BEC in the lower dimensions \cite{2D_liquid,2D_liquid1, 2D_liquid2,2D_liquid3, Gajda19}.

% SOC

The spin-orbit coupling(SOC), the coupling between the spin angular momentum of a particle with its center of mass motion, is present in atomic electrons, electrons in some solids\cite{spintronics}. Based on the symmetry of interaction, we have two types of 2D SOC in solid state systems such as Rashba\cite{Rashba} and Dresselhaus SOC\cite{Dresselhaus}. The SOC in 2D solid produces the topological spin Hall effect. It has been observed in neutral ultracold atoms with the help of atom-light interaction\cite{SOC_BEC_exp, SOC_BEC_rev}. 
The study of BEC in the system of spin-orbit coupled Bose gas became very important after the discovery of Lin and et. al.\cite{SOC_BEC_exp}.
Chiquillo has theoretically addressed the liquid phase of two spices Bose atoms with Rabi-Coupling\cite{2D_liquid_Rabi}. 

%ground state Theory

Theoretically, the system can be addressed in two different approaches such as the mean-field Gross--Pitaevskii (GP) equation, which is involved in the solution of time dependent non-linear coupled differential equations and another one is the Monte-Carlo integration with some suitable microscopic wave function\cite{MC}.  GP equations have been solved for isotropic droplets of two spices atoms for different kinds of interactions\cite{Adhikari2018} and anisotropic dipolar system\cite{GP_dipole}. Here we have applied the GP equation technique as it is most popular and it has the advantage to get the collective excitation using Bogoliubov approach.

% excitations
%The elementary excitation of conventional superfluid of liquid $^4$He is the phonon-roton mode\cite{Feynman_Roton}.
The elementary excitation in the uniform BEC of infinite extent is easy to study using Bogoliubov theory. In this weak interacting system of ultracold dilute gas of Bose atoms in the condensate,  the low energy collective excitation is the well known phonon-mode\cite{Pethick_book}. On the other hand, the excitation in the nonuniform confined state is discrete due to the finite size of the system which has been studied theoretically\cite{trap_exc1} and experimentally \cite{exc_disc_exp}. 
The collective excitation of uniform BEC of atoms with SOC has been studied theoretically\cite{Stringari2012} and experimentally\cite{coll_exc_SOC_BEC}. They got the phonon-roton mode of excitation. 

The liquid droplets are finite in size, but it has completely different properties than confined BEC. The density inside the droplet is constant except near the surface. To study the collective excitation inside the liquid droplets, we have considered the sufficiently large droplets to avoid the surface effect. To find the Bogoliubov spectra, we have used the constant background density, which is obtained by solving the GP equations. So the excitation is equivalently the collective excitation of bulk liquid of ultra-cold atoms.

%

% we did
In our study, we have considered the two dimensional BEC of a mixture of atoms of two internal degrees of freedom of the same element.  The short range delta function repulsive potential between the atoms of same species and attractive potential between atoms of different species has been considered.
 We have considered two types of SOC, such as the combination of  Rashba and Dreshelhaus SOC with different strength and the Raman SOC.

%we get
%In our study we have seen that the roton type of excitations is only possible in the strong interacting regime in the absence of SOC. In presence of SOC the phonon-roton mode is present even in the weak coupling regime.

\section{Model and method of calculations}

The short range interaction potential between two atoms can be written as $V(\vec{r}_1,\vec{r}_2) = g_{ij} \delta(\vec{r}_1-\vec{r}_2)$, $g_{ij}$ is the strength of the interaction of atoms of $i$th and $j$th species, can be expressed as $g_{ij}=\frac{4\pi \hbar^2 a_{ij}}{m}$, where $m$ is the mass of each atom, $a_{ij}$ is the scattering length, which can be controlled by magnetic Feshbach resonance.
Near the phase transition the mean-field GP equations are not sufficient to study the nature of the condensate, we need to consider self-repulsive beyond-mean-field, Lee-Huang-Yang (LHY) term (quantum fluctuation term)\cite{LHY}. 
The  LHY interaction in the 2D system takes the logarithmic form\cite{2D_liquid}, and the corresponding nonlinear coupled GP equations of two component BEC in presence of Rashba and Dreshelhaus SOC can be written as\cite{RD_SOC}
%\small
%\begin{widetext}
\begin{eqnarray}
%     i\frac{\partial\psi_1 }{\partial t}=\left[-\frac{\nabla ^2}{2} + g(|\psi_1|^2-|\psi_2|^2) + \frac{g^2}{4\pi}(|\psi_1|^2+|\psi_2|^2) ln (|\psi_1|^2+|\psi_2|^2)  \right ]\psi_1+(\lambda_RD-i\lambda_DD^*)\psi_2.\\
 %         i\frac{\partial\psi_2 }{\partial t}=\left[-\frac{\nabla ^2}{2} + g(|\psi_1|^2-|\psi_2|^2) + \frac{g^2}{4\pi}(|\psi_1|^2+|\psi_2|^2) ln (|\psi_1|^2+|\psi_2|^2)  \right ]\psi_2-(\lambda_RD^*+i\lambda_DD)\psi_1.\\
  i\frac{\partial\Psi}{\partial t} &=& -\frac{\nabla ^2}{2} \Psi+ g(|\psi_1|^2-|\psi_2|^2) \sigma_z \Psi \\%nonumber \\
                                   &+& \frac{g^2}{4\pi} \rho \;ln(\rho ) \Psi  +\lambda_R \vec{\sigma}\cdot \hat{z}\times \vec{p} \Psi + \lambda_D \vec{\sigma}\cdot \vec{p} \Psi  \nonumber
\end{eqnarray}
%\end{widetext}
%\normalsize
All the quantities are expressed in sutable natural units of the system\cite{Adhikari2018}.
 First term of the right hand side is the kinetic energy term, second term is the interaction potential term, third term is the quantum  fluctuation term, fourth term is the Rashba SOC term with coupling strength $\lambda_R$, the last term is the Dresselhaus term with coupling strength $\lambda_D$.  $\Psi=(\psi_1 \; \psi_2)^T$ is the two components wave function and ($\sigma_x$, $\sigma_y$, $\sigma_z$) are the Pauli spin matrices.   Here we have considered $g_{11}=g_{22}=-g_{12} = g$.
The normalization of the wave function is given by
\begin{eqnarray}
\int \left ( |\psi_1|^2+|\psi_2|^2\right ) d^2r =\int \rho(\vec{r}) d^2r = N, %\nonumber
\end{eqnarray}
where $N$ is the total number of particles in the system, $\rho$ is the density of the condensate.
 We have used Crank Nicolson method to solve the nonlinears coupled equations. Alternating direction implicit method\cite{ADC_method} has been used to tackle the 2D derivatives. 
 
%figure 1
\begin{figure}[htbb]
  \begin{center}
        \includegraphics[width=0.48\textwidth]{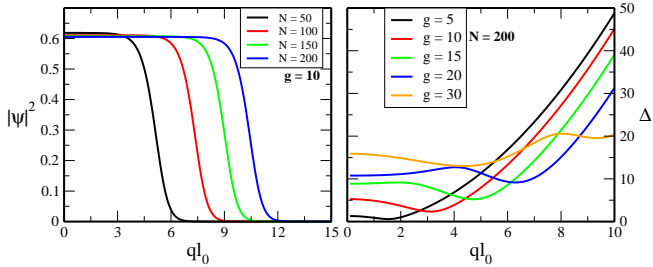}                

\caption{{\bf Circular liquid film in absence of SOC:} The nature of the ground state as well as excited state in the absence of SOC ($\lambda_R=\lambda_D=0$). Left panel: The density of the condensate as a function of distance from the centre for different number of particles for a fixed value of $g$(=10.0). Right panel: The collective excitations of the liquid film for different values of $g$ as indicated in side the figure.}
    \label{2D_LHY}
  \end{center}
\end{figure}

%figure 2
\begin{figure}
  \begin{center}
     \includegraphics[width=2.5cm]{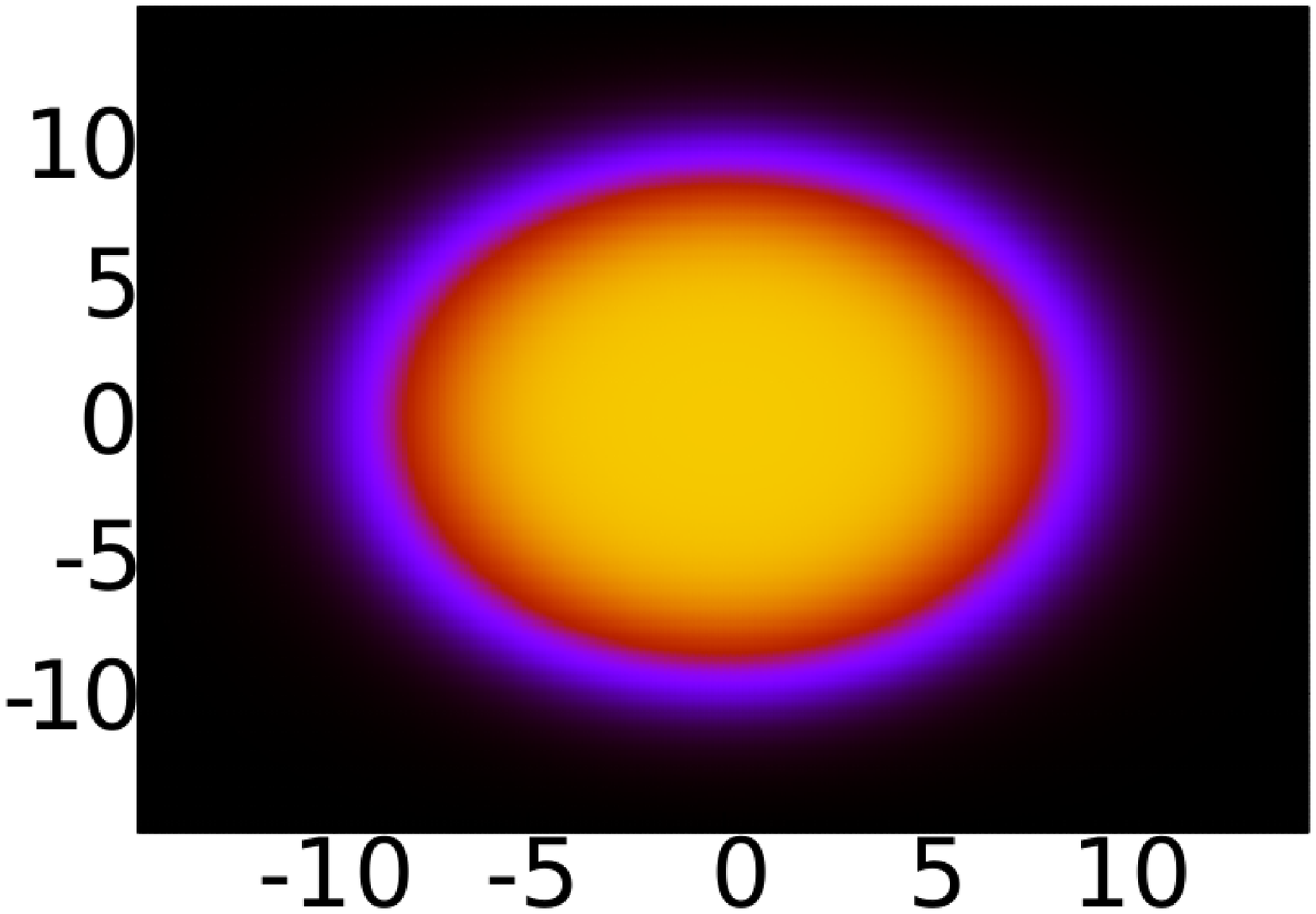} 
      \includegraphics[width=2.5cm]{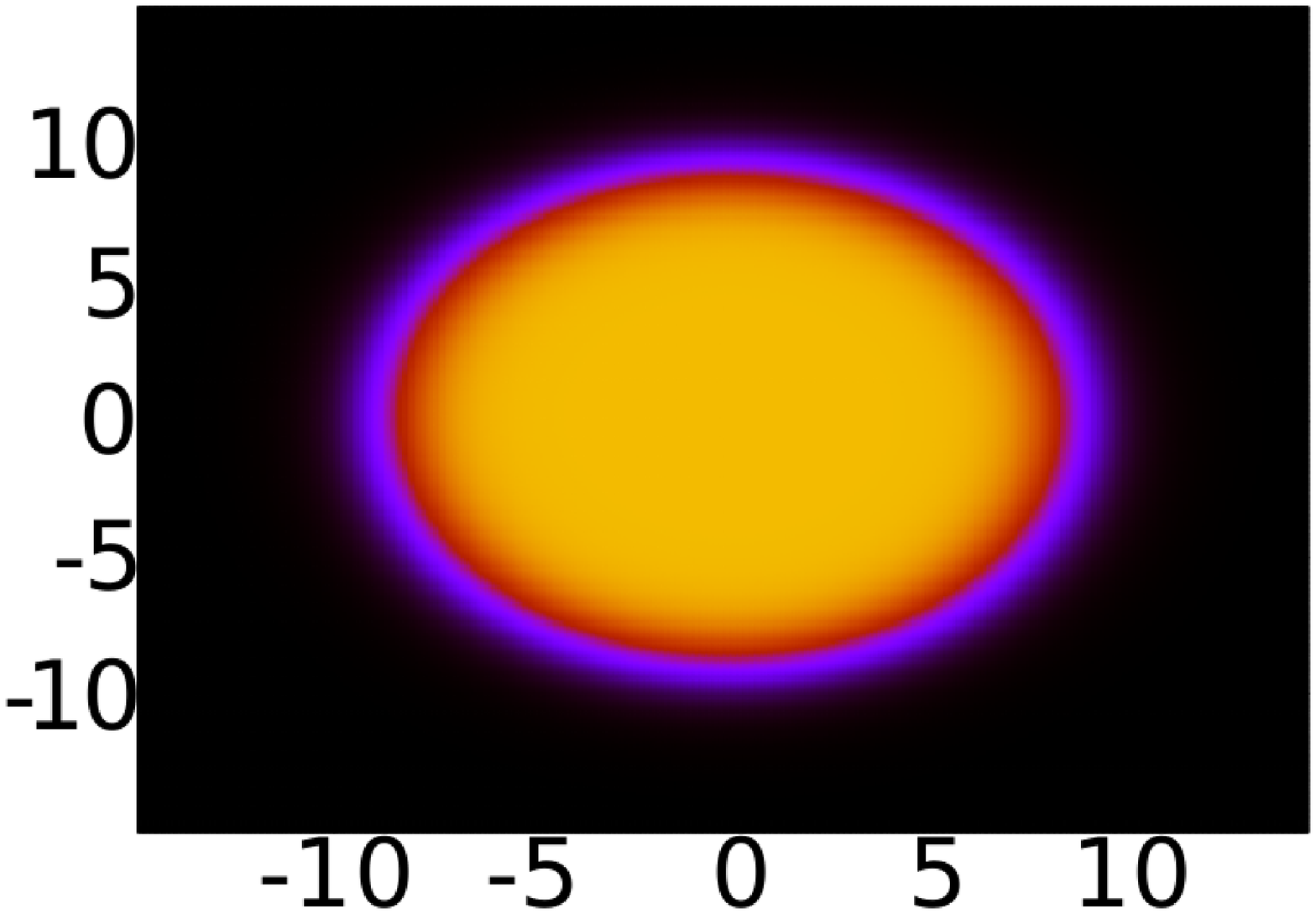}
       \includegraphics[width=2.5cm]{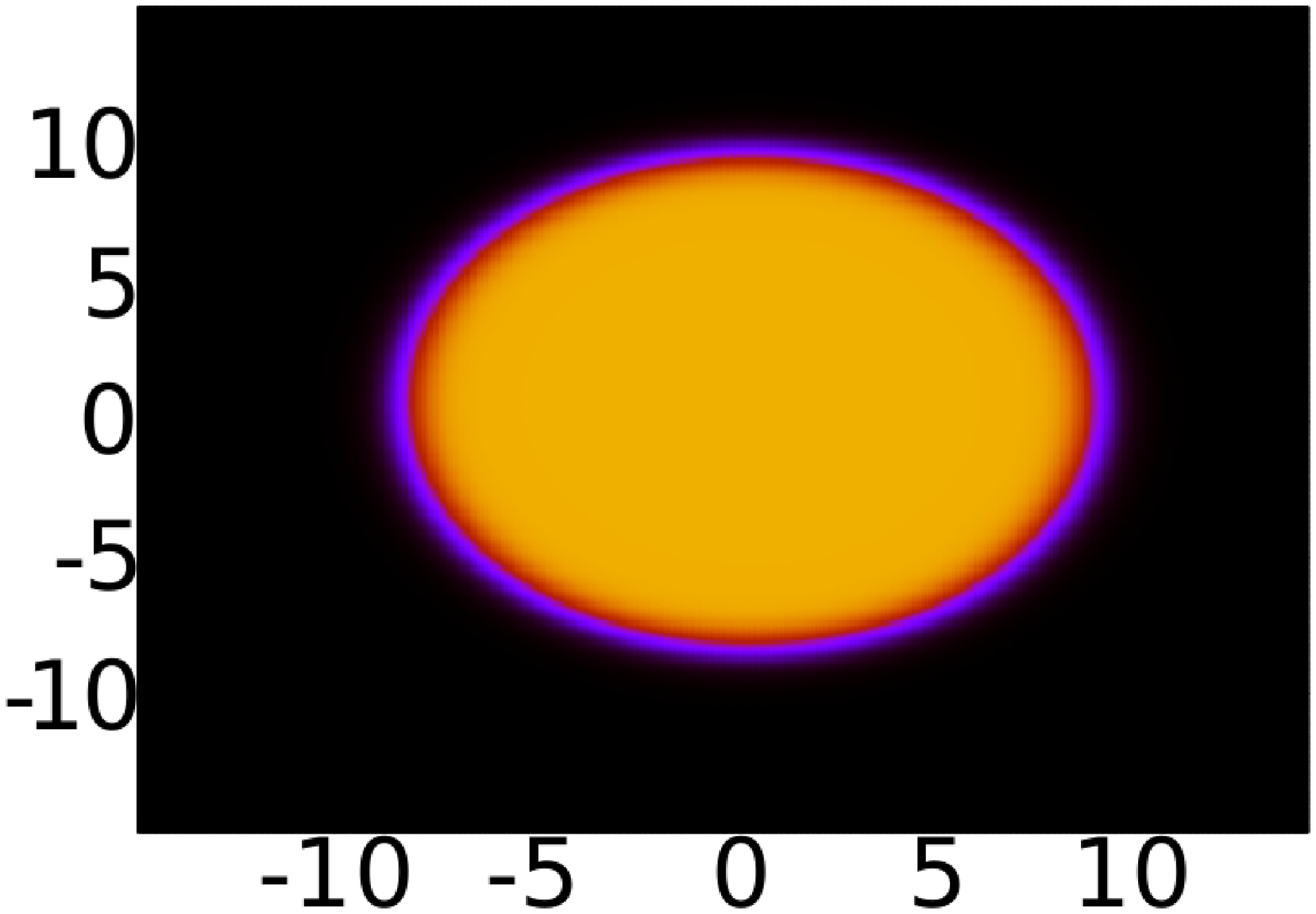}
       \includegraphics[height=1.75cm ]{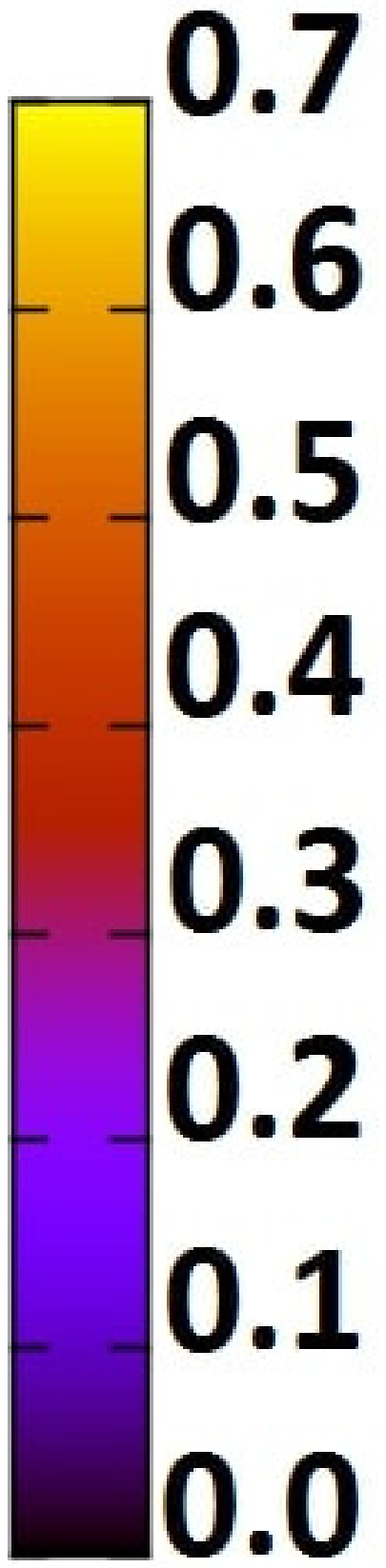}
       
       \includegraphics[width=2.5cm]{d0r2g10.eps}
       \includegraphics[width=2.5cm]{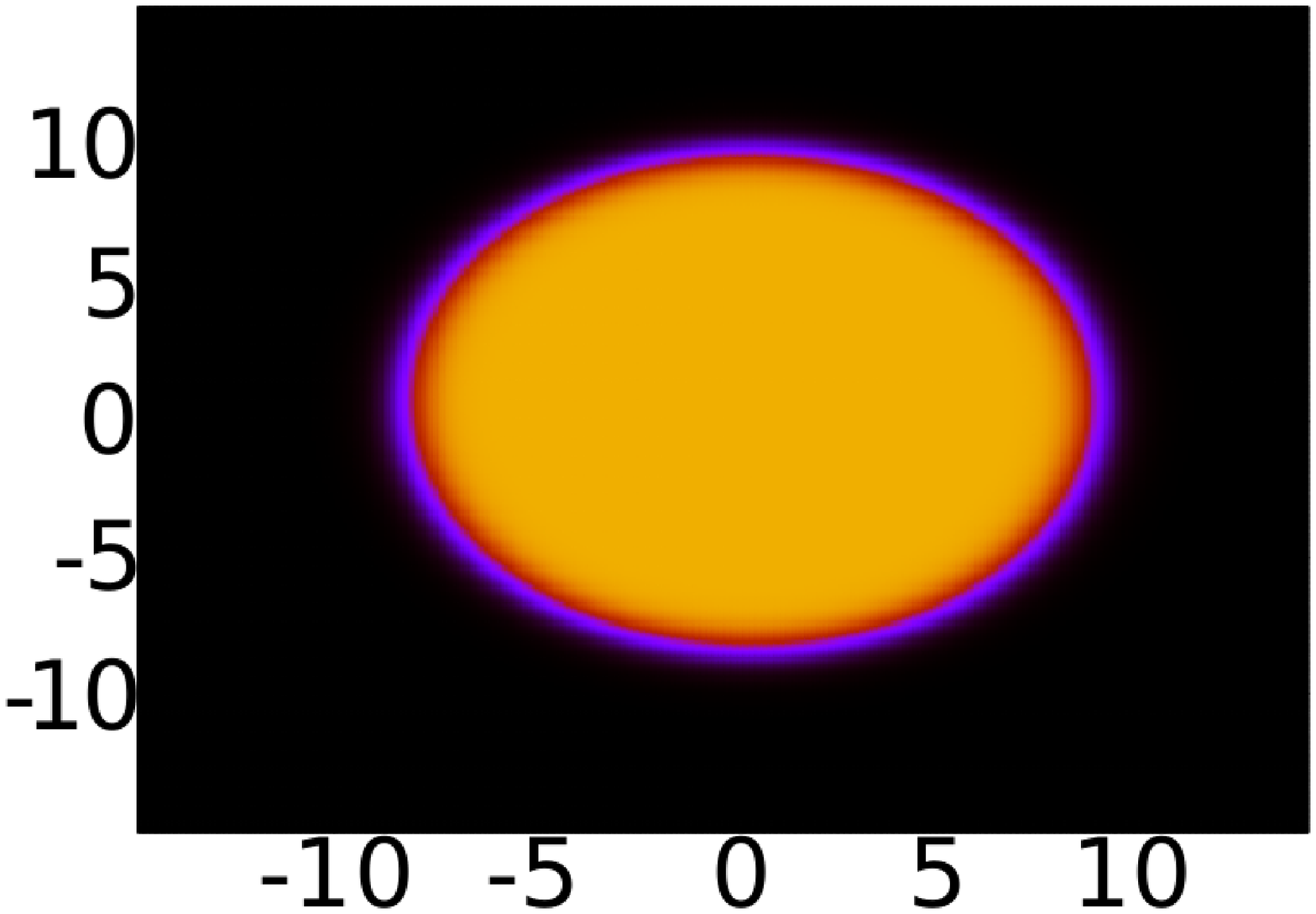}
       \includegraphics[width=2.5cm]{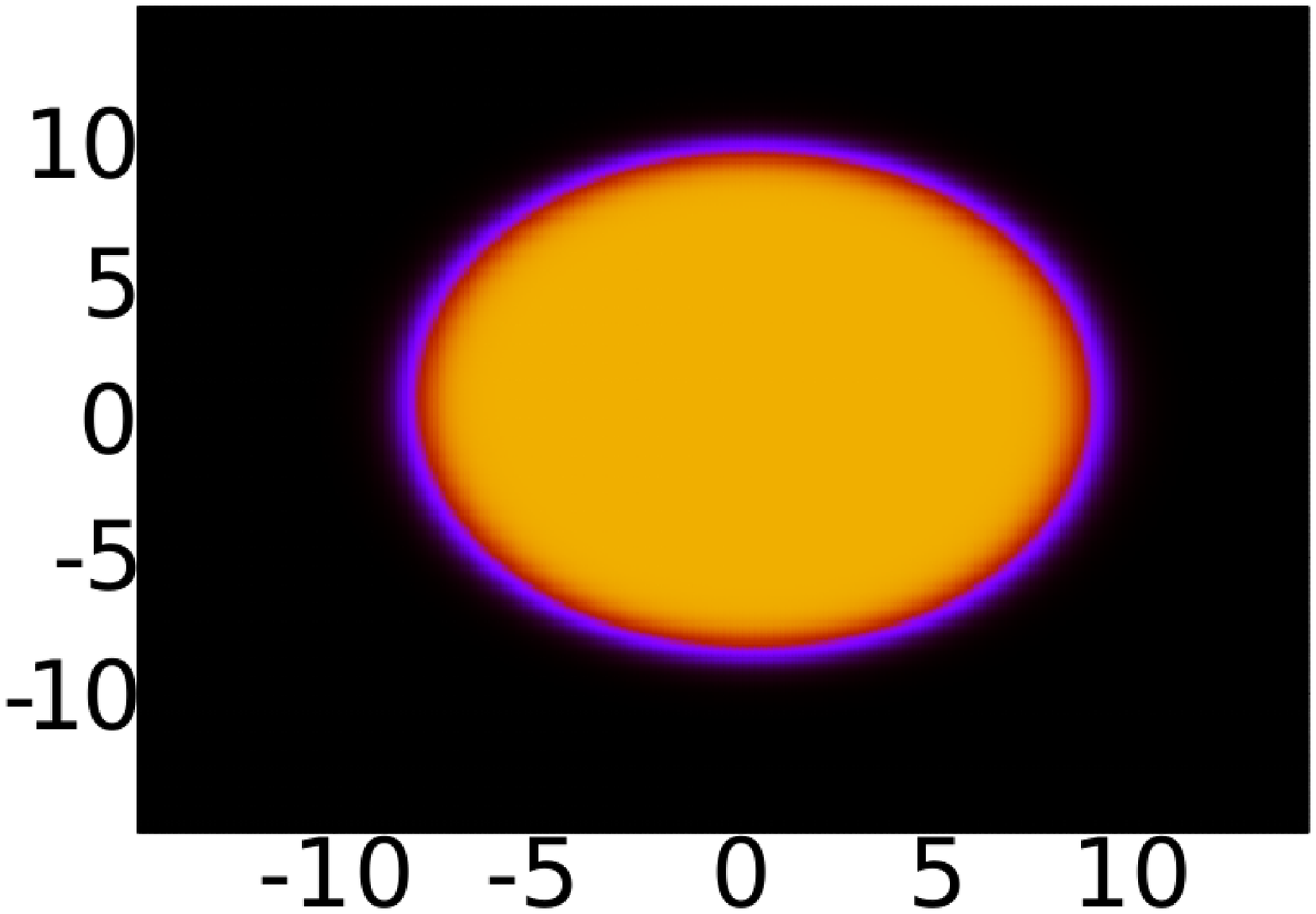}
              \includegraphics[height=1.75cm]{colorbar.eps} 
\caption{ Ground state density for various strength of Rashba SOC and two body interaction  for 150 number of atoms. 
{\bf 1st Row:}  $g = 3, 5, 10$ respectively, for $\lambda_R = 0.2$. 
{\bf 2nd Row:}  $\lambda_R = 0.2, 0.3, 0.5$ respectively for $g = 10$.}
    \label{density_Rashba}
    
        \includegraphics[width=2.5cm]{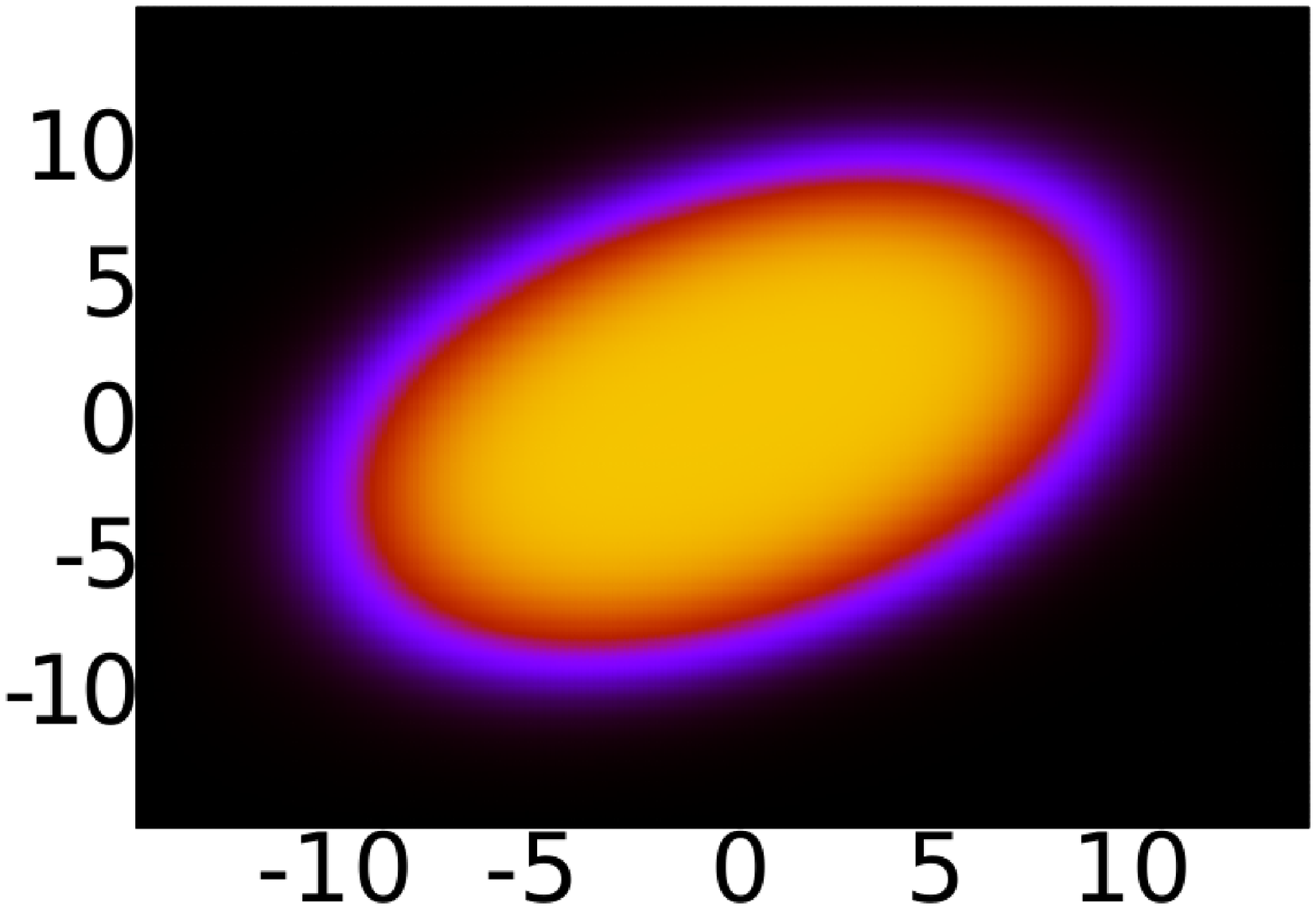} 
       \includegraphics[width=2.5cm]{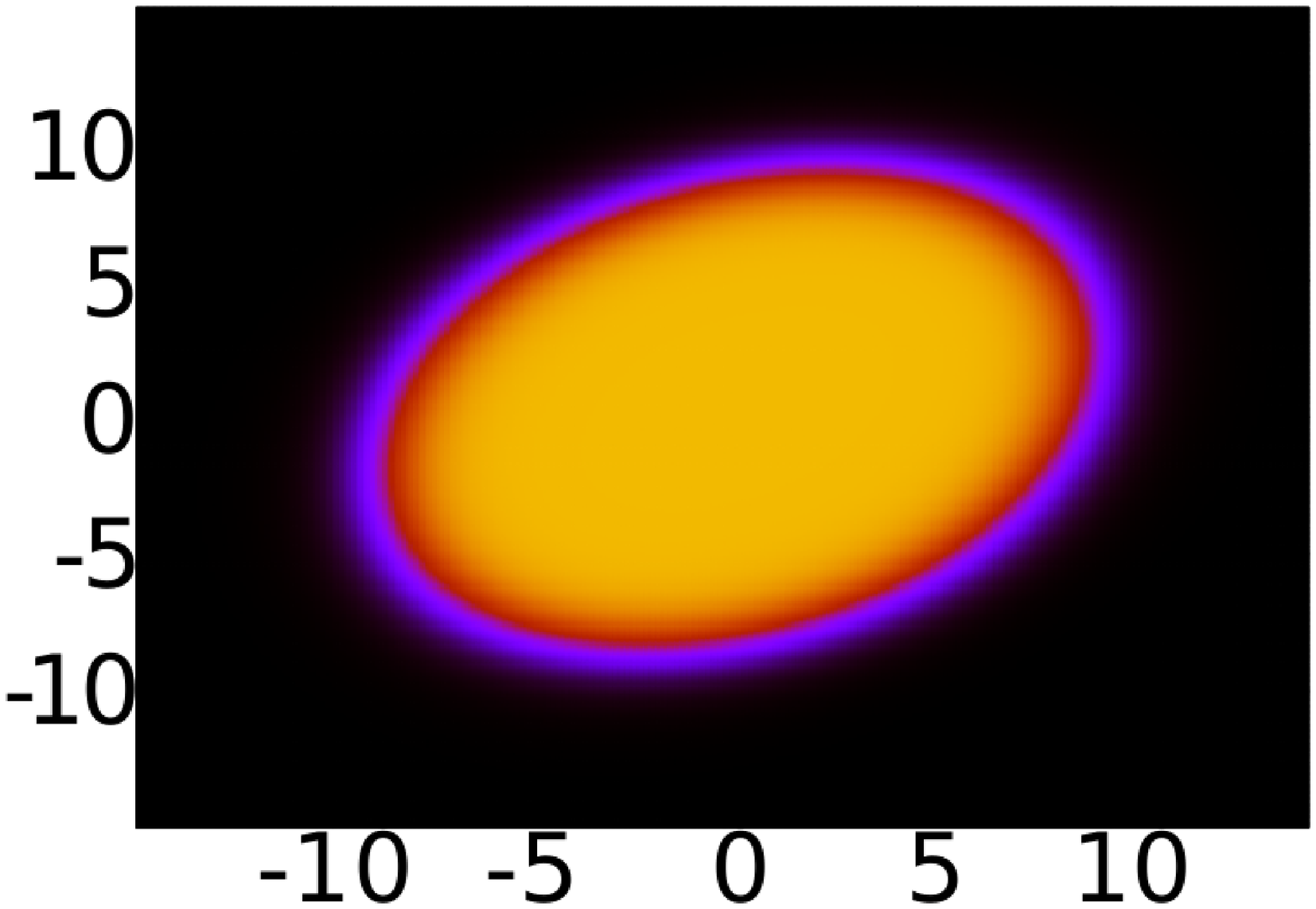}
       \includegraphics[width=2.5cm]{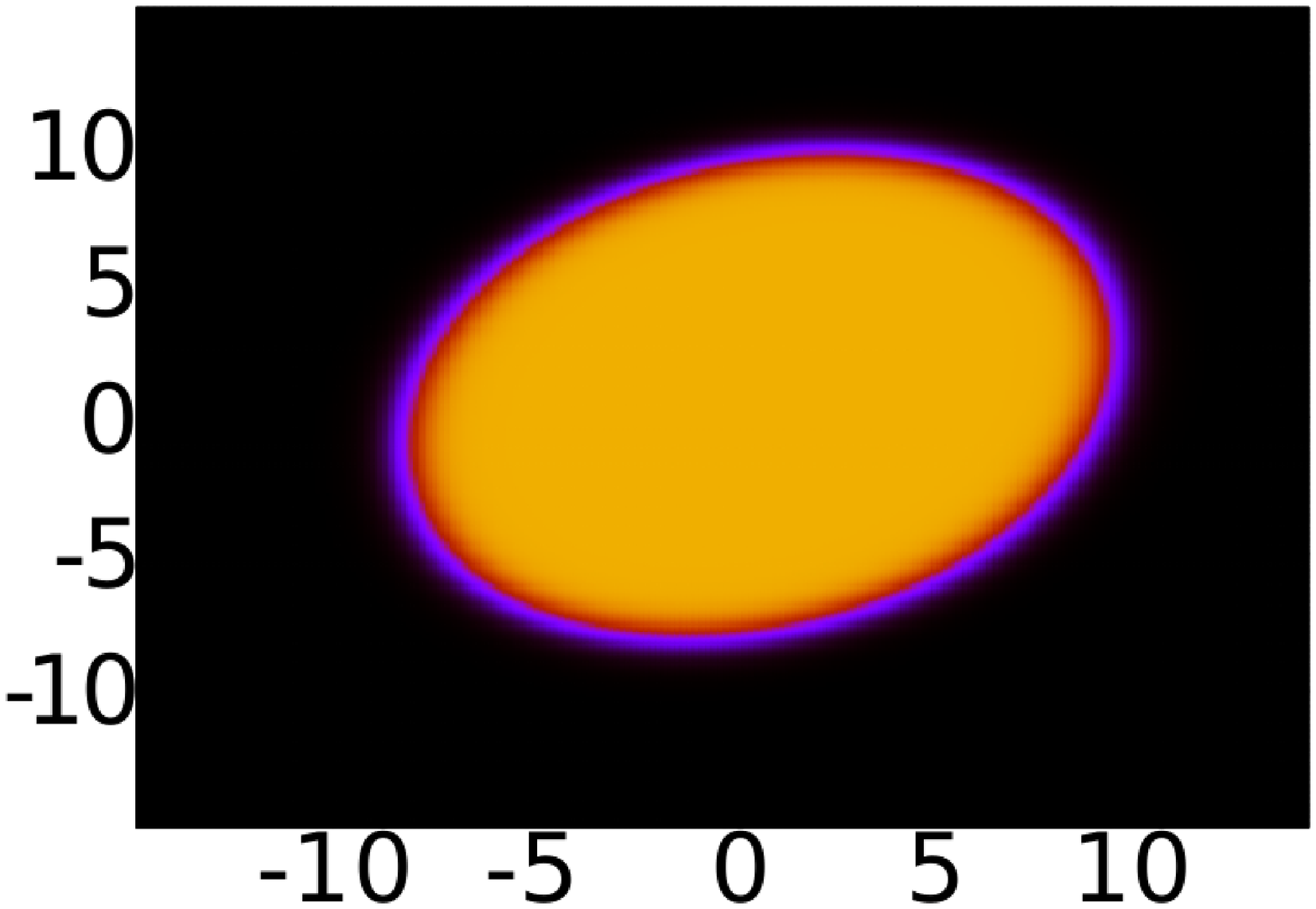}
       \includegraphics[height=1.75cm]{colorbar.eps}
       
       \includegraphics[width=2.5cm]{d2r1g10.eps}
       \includegraphics[width=2.5cm]{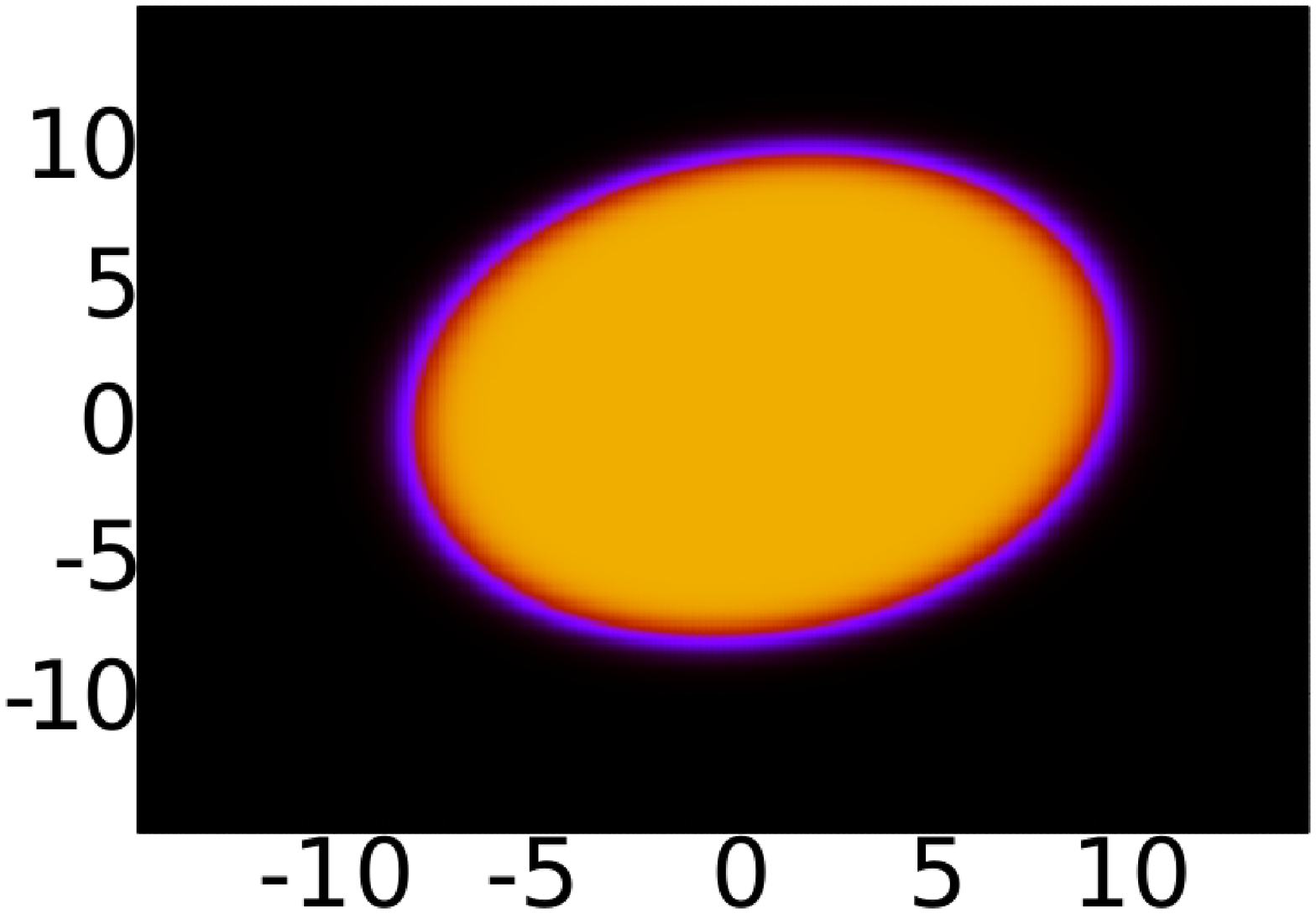}
       \includegraphics[width=2.5cm]{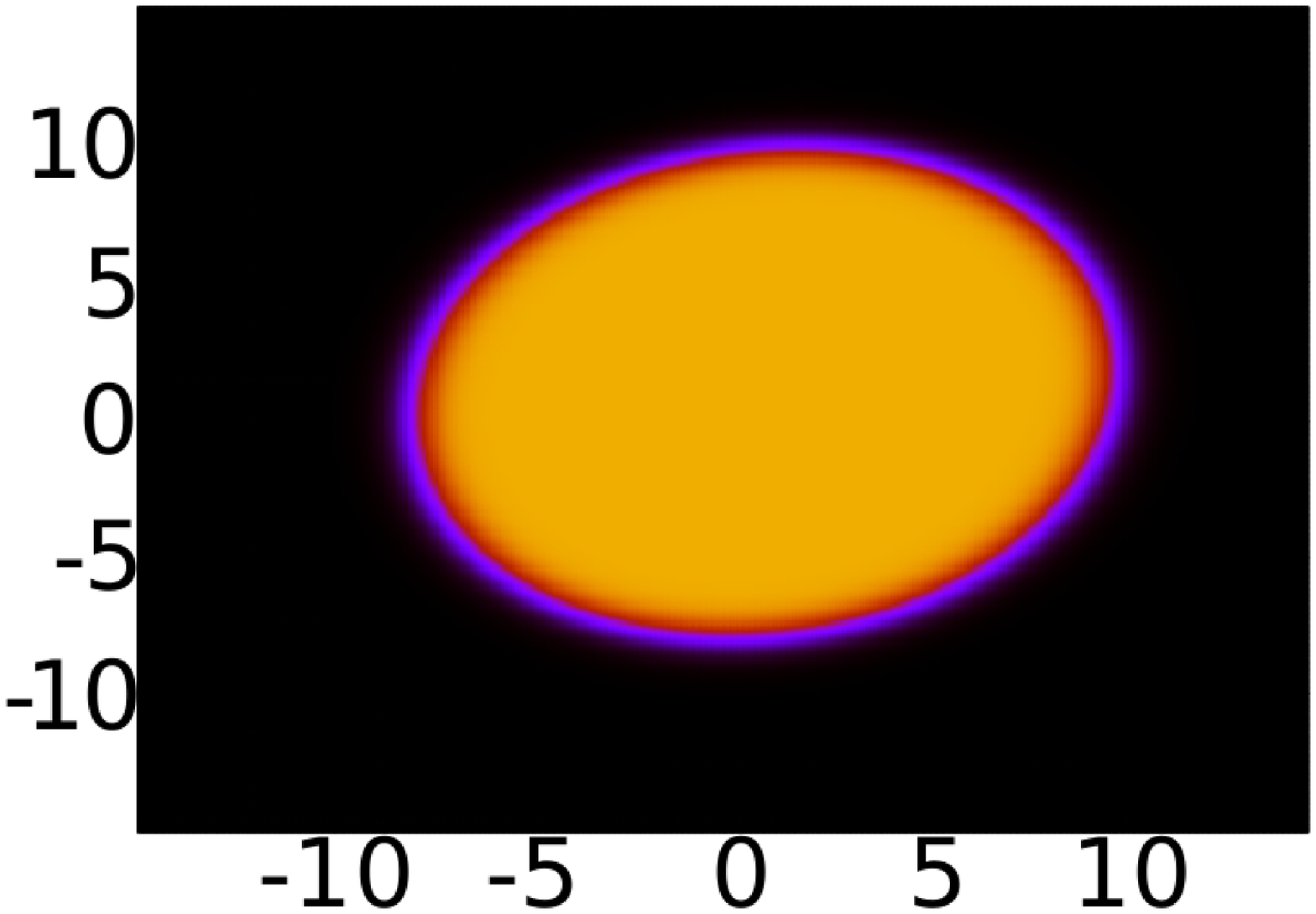}
              \includegraphics[height=1.75cm]{colorbar.eps} 
\caption{ Ground state density for various strength of Dresselhaus SOC  and two body interaction  for 150 number of atoms at fixed Rashba interaction $\lambda_R=1.0$. 
{\bf 1st Row:}  $g = 3, 5, 10$ respectively, for $\lambda_D = 0.2$. 
{\bf 2nd Row:}  $\lambda_D = 0.2, 0.3, 0.5$ respectively for $g = 10$.}
    \label{density_RD}
  \end{center}
\end{figure}

% figure 3
\begin{figure}
  \begin{center}
    \includegraphics[width=0.5\textwidth]{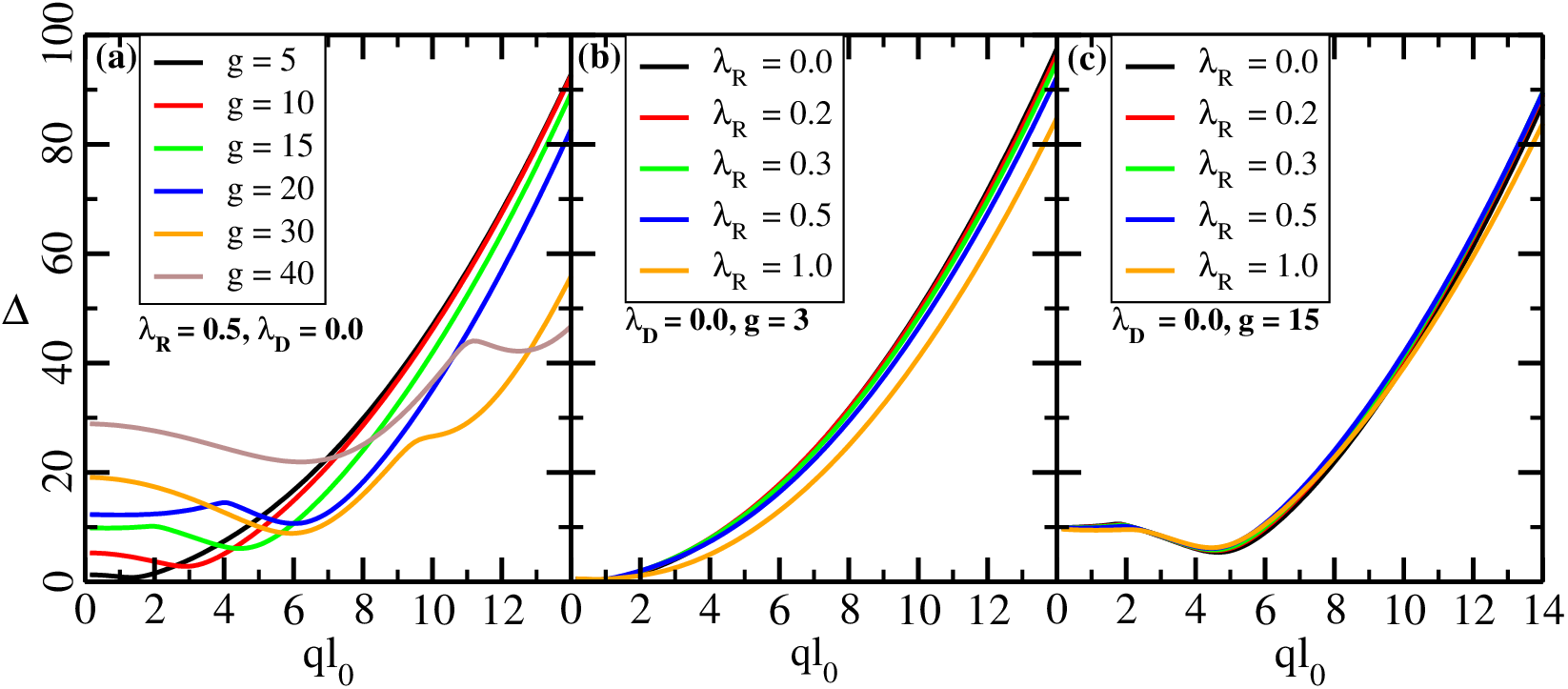}
    \caption{Rashba SOC: Collective excitation of the condensate under the Rashba SOC. (a) Variation of excitations with $g$ for fixed Rashba SOC. (b),(c) Variation with Rashba SOC for fixed $g$. }
    \label{exc_RSOC}    
      %\includegraphics[width=0.42\textwidth]{dress_exc.eps}  
      %\caption{Dresselhaus}
          \includegraphics[width=0.5\textwidth]{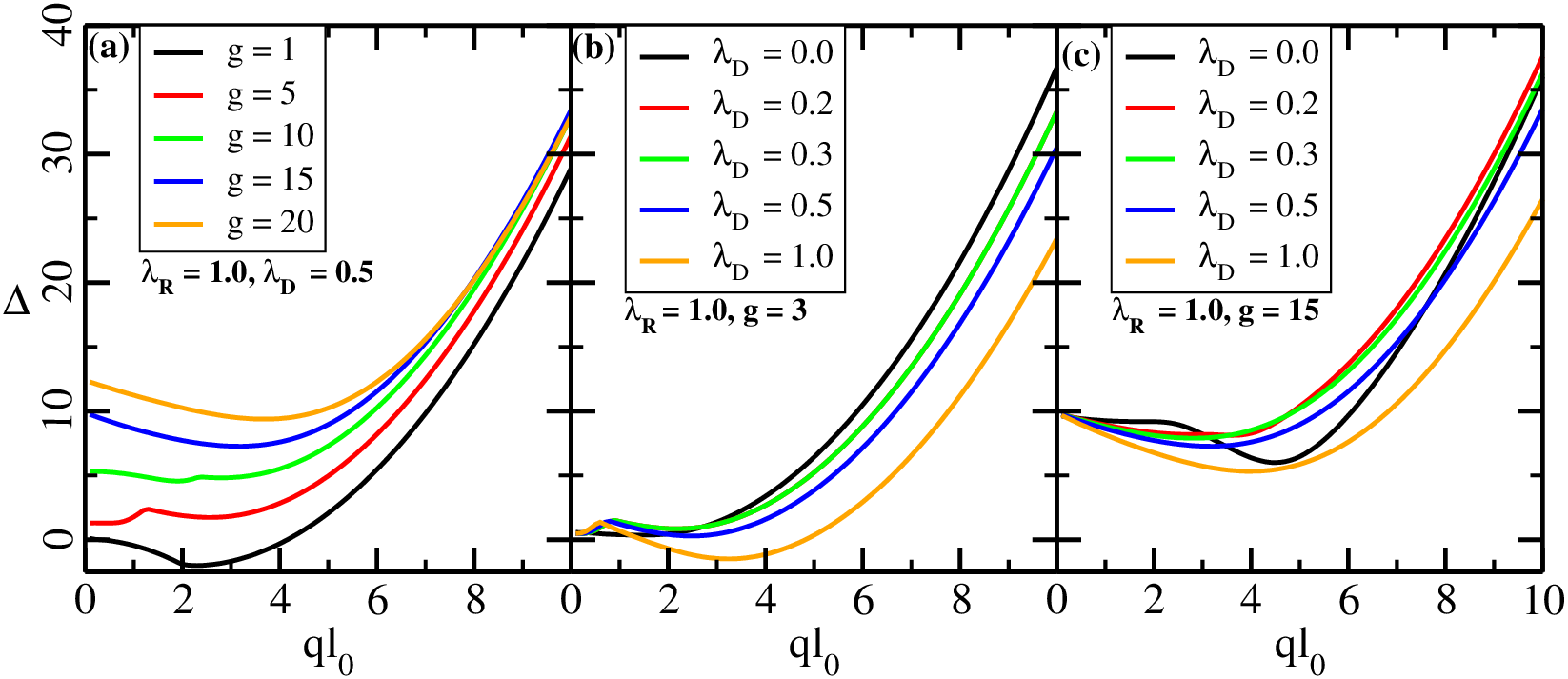}
  \caption{Rashba and Dresselhaus SOC combination: Collective excitation of film of atoms with combined Rashba and Dresselhaus SOC. }
    \label{exc_RDSOC}
   \end{center}
\end{figure} 

 Li et al.\cite{RD_SOC} has established the region of mixed mode in the parameter space. We have calculated the ground state density and excitation in the mixed mode region, where the two species superpose uniformly on eachother\cite{MM_SOC}. The ground state density has been plotted in the left panel of Fig. \ref{2D_LHY} as a function of distance from the center of the condensate for the system without SOC. Its shows that the density does not depend on the number of particles. If we increase the number of particles the size of the film will increase. 
The ground state density for different combination of Rashba and Dresselhaus SOC has been shown in Fig. \ref{density_Rashba} and in Fig. \ref{density_RD}.

\subsection{Collective excitation} %Bogoliubov excitation Energy}

After getting the ground state wave function, we have applied Bogoliubov theory of excitation over the ground state. We chose the excited state wave function as a perturbation over the ground state, $\psi_j^{\mbox{exc}} = \psi_j+\delta \psi_j$. The perturbation part of the wave function can be written as $\delta \psi_j = U_j exp(i\vec{q}\cdot  \vec{r} -i\omega t) -V_j^* exp (-i\vec{q}\cdot  \vec{r} +i\omega t)$, where  $\{U_j, V_j\}$ are amplitude of the excitation, $\vec{q}$ is quasi-momentum of the excitation and $\omega$ is the excitation energy. If we put these in our GP equations we will find the equations for $\{U_j, V_j\}$, in the first order approximation of $\{\delta\psi_i\}$
\begin{widetext}
\small 
  \begin{eqnarray}
\left(
    \begin{tabular}{c c c c}
      $H_{1}$ & $X+A {\psi}_1 {\psi}_2^*$ & $B {\psi}_1^2$&$ A{\psi}_1 {\psi}_2$\\
       $X^*+A {\psi}_1^* {\psi}_2$ & $H_2$ & $A {\psi}_1 {\psi}_2$ &  $B {\psi}_2^2$\\
       $-B{\psi}_1^{*2}$ & $-A{\psi}_1^* {\psi}_2^*$ &$-H_1$ & $-X^*-A {\psi}_1^* {\psi}_2$\\
       $-A {\psi}_1^* {\psi}_2^*$ & $-B {\psi}_2^{*2}$ & $-X-A {\psi}_1 {\psi}_2^*$ & $-H_2$\\
    \end{tabular}
\right)
\left(
\begin{tabular}{c}
$U_1$ \\$U_2$\\$V_1$\\$V_2$ 
\end{tabular}
\right) = \omega \left(
\begin{tabular}{c}
$U_1$ \\$U_2$\\$V_1$\\$V_2$ 
\end{tabular}
\right) 
\end{eqnarray}
\normalsize
\end{widetext}

where 
%\small 
\begin{eqnarray}
H_1 &=&  \frac{q^2} {2}+ 2g |\psi_1|^2- g |\psi_2|^2+\frac{g^2}{4\pi}(2|\psi_1|^2+|\psi_2|^2) ln (\rho) \nonumber \\
H_2 &=&  \frac{q^2} {2}+ 2g|\psi_2|^2- g |\psi_1|^2+\frac{g^2}{4\pi}(2|\psi_2|^2+|\psi_1|^2) ln (\rho) \nonumber \\
A &=& -g +\frac{g^2}{4\pi} ln (\rho) \nonumber \\
B &=& g +\frac{g^2}{4\pi} ln (\rho) \nonumber \\
X &=& \lambda_R(q_y+iq_x)+\lambda_D(q_x+iq_y) \nonumber
\end{eqnarray}

\normalsize

The energy eigen value is obtained by digonalizing the matrix. The results have been shown in the right panel of FIG \ref{2D_LHY} for the system without SOC and in FIG \ref{exc_RDSOC} and FIG \ref{exc_RSOC} for the SOC. The discussion of the results have been done in the result and discussion section.

\subsection{Raman SOC}

After the discovery of the SOC-BEC in 2011\cite{SOC_BEC_exp}, we have another form of spin-orbit coupling, the Raman SOC. The coupled GP equations for the system with $g_{1,1}=g_{2,2}=-g_{1,2}=g$, can be written as

\begin{figure}
  \begin{center}
       \includegraphics[width=2.5cm]{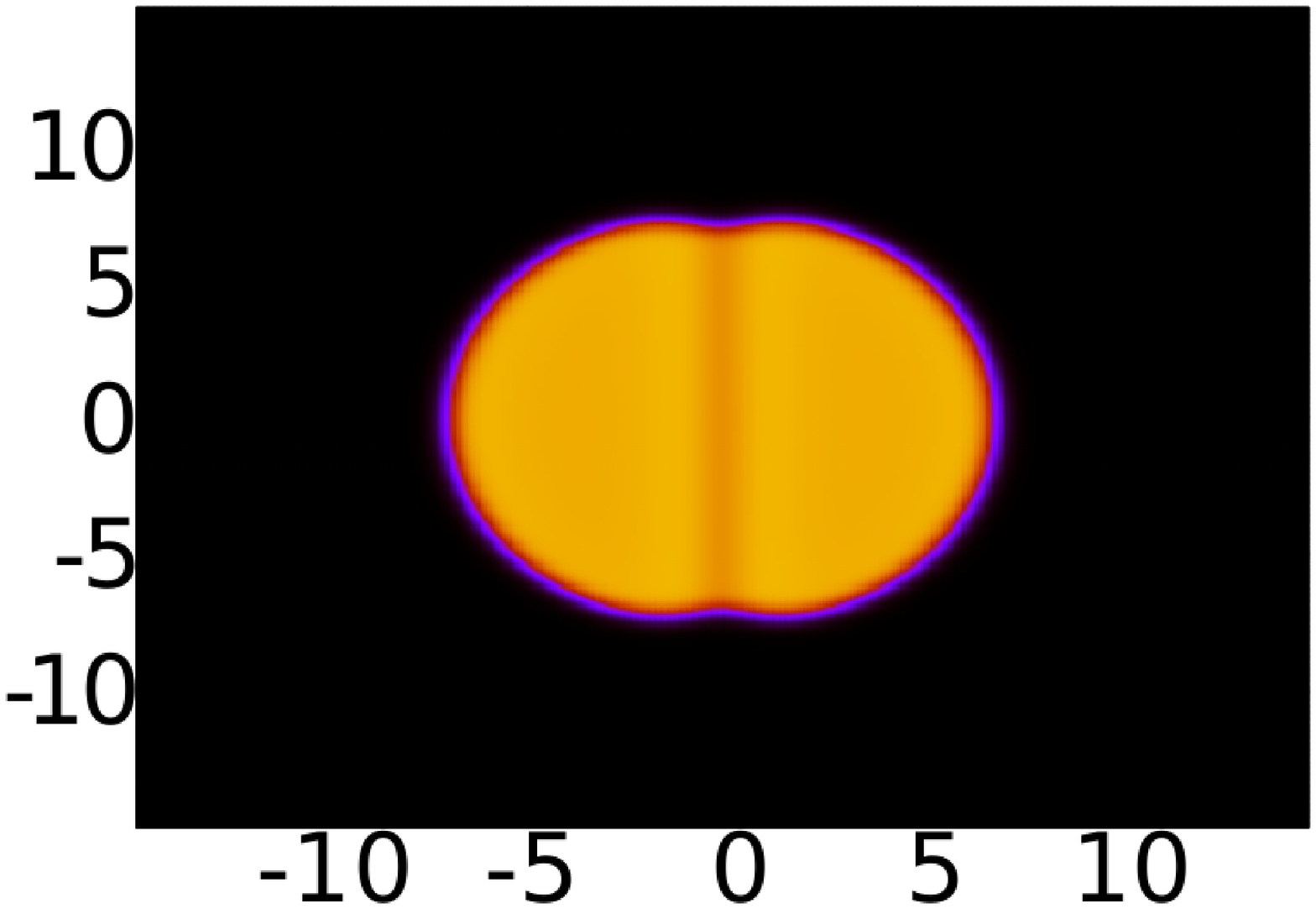} 
       \includegraphics[width=2.5cm]{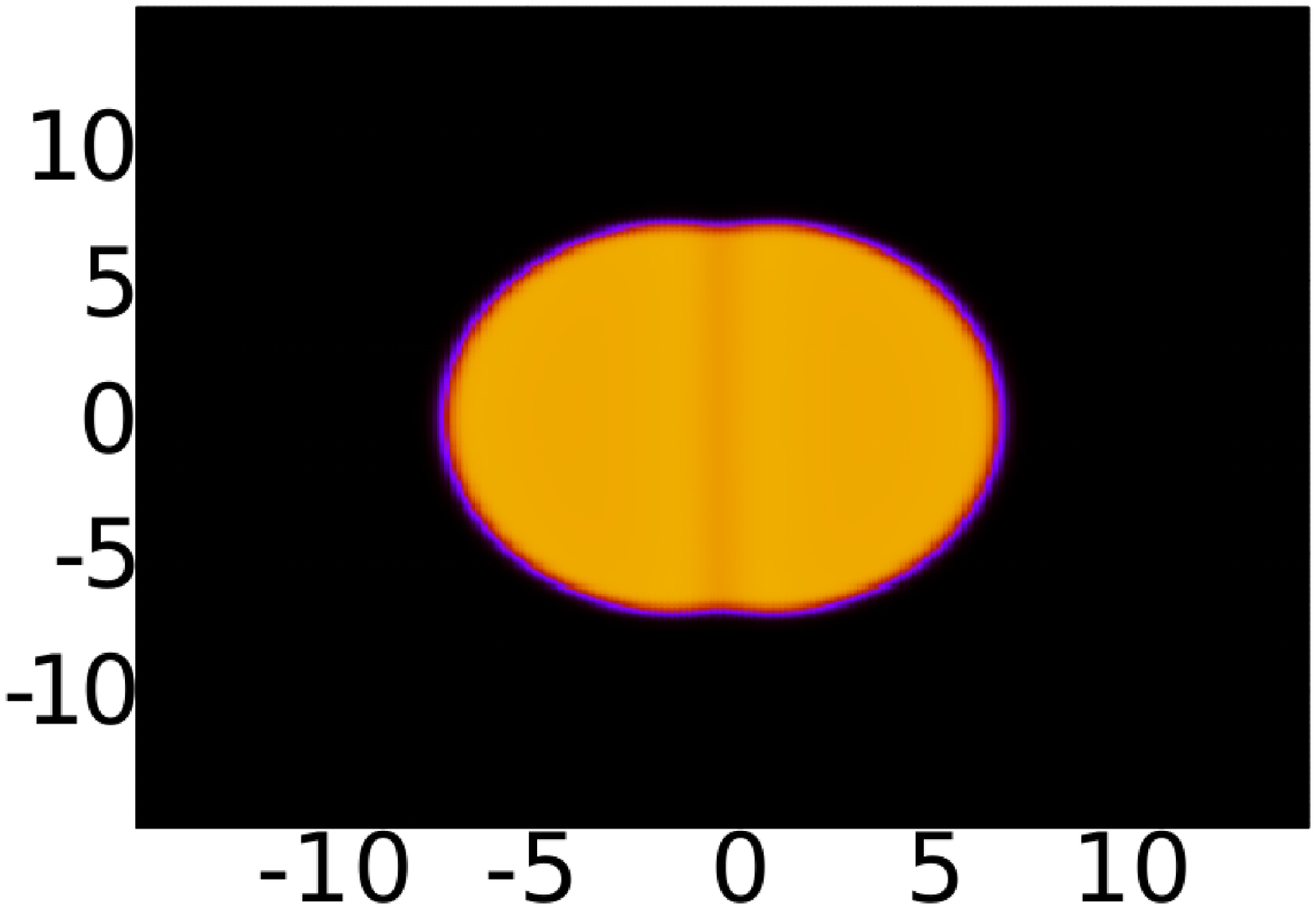}
       \includegraphics[width=2.5cm]{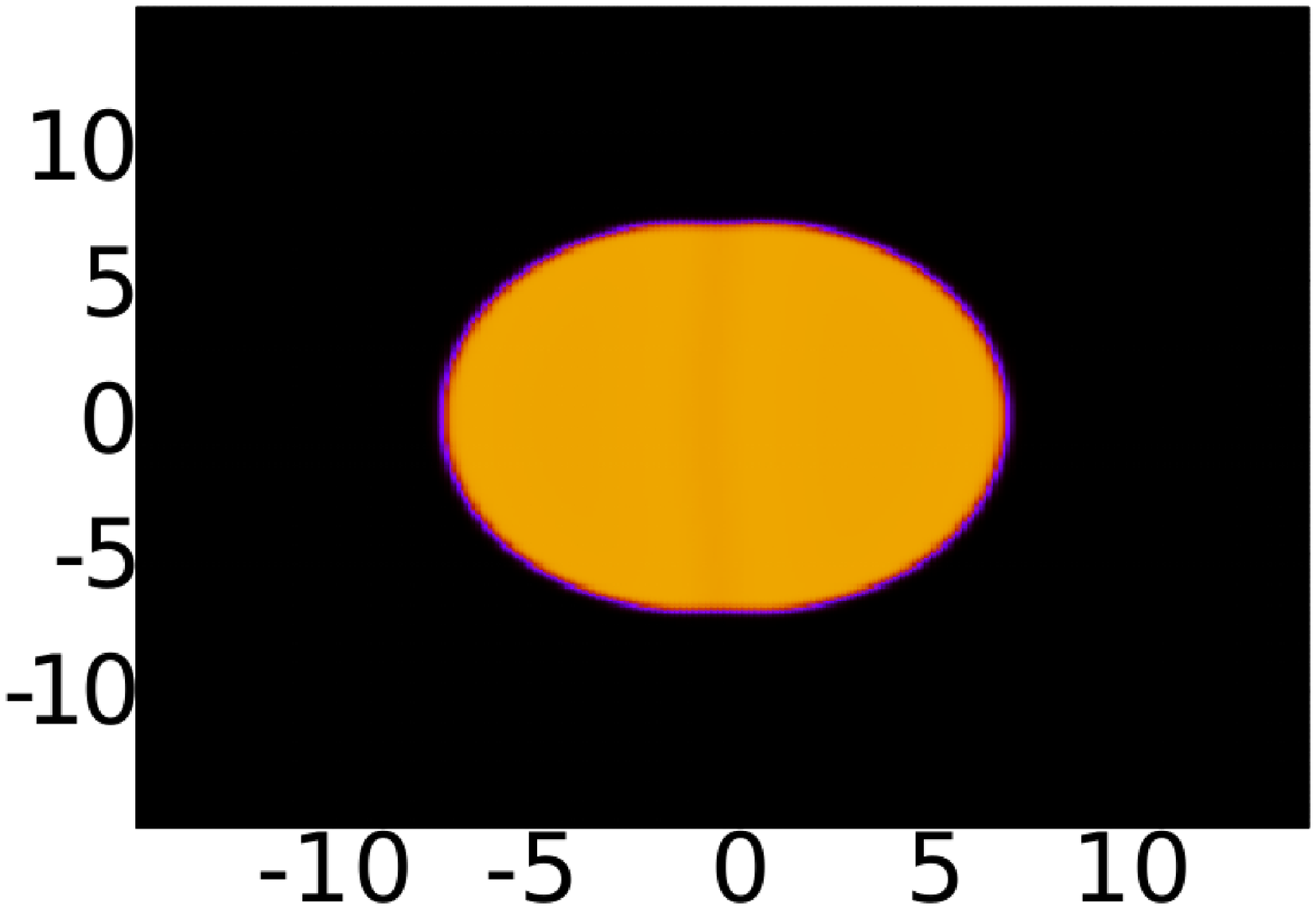}
       \includegraphics[height=1.75cm]{colorbar.eps}\\
       \includegraphics[width=2.5cm]{k1om1g50.eps}
       \includegraphics[width=2.5cm]{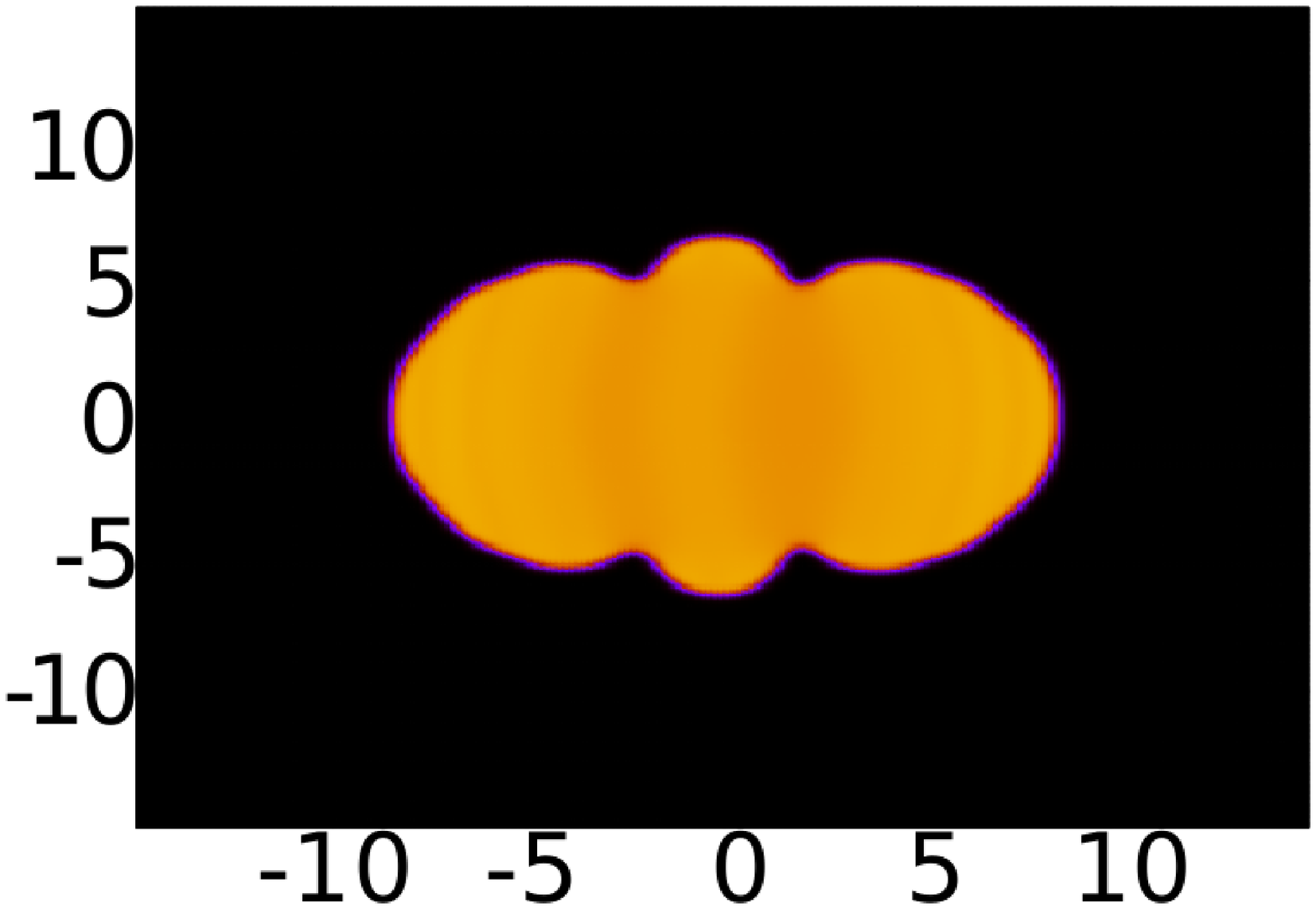}
       \includegraphics[width=2.5cm]{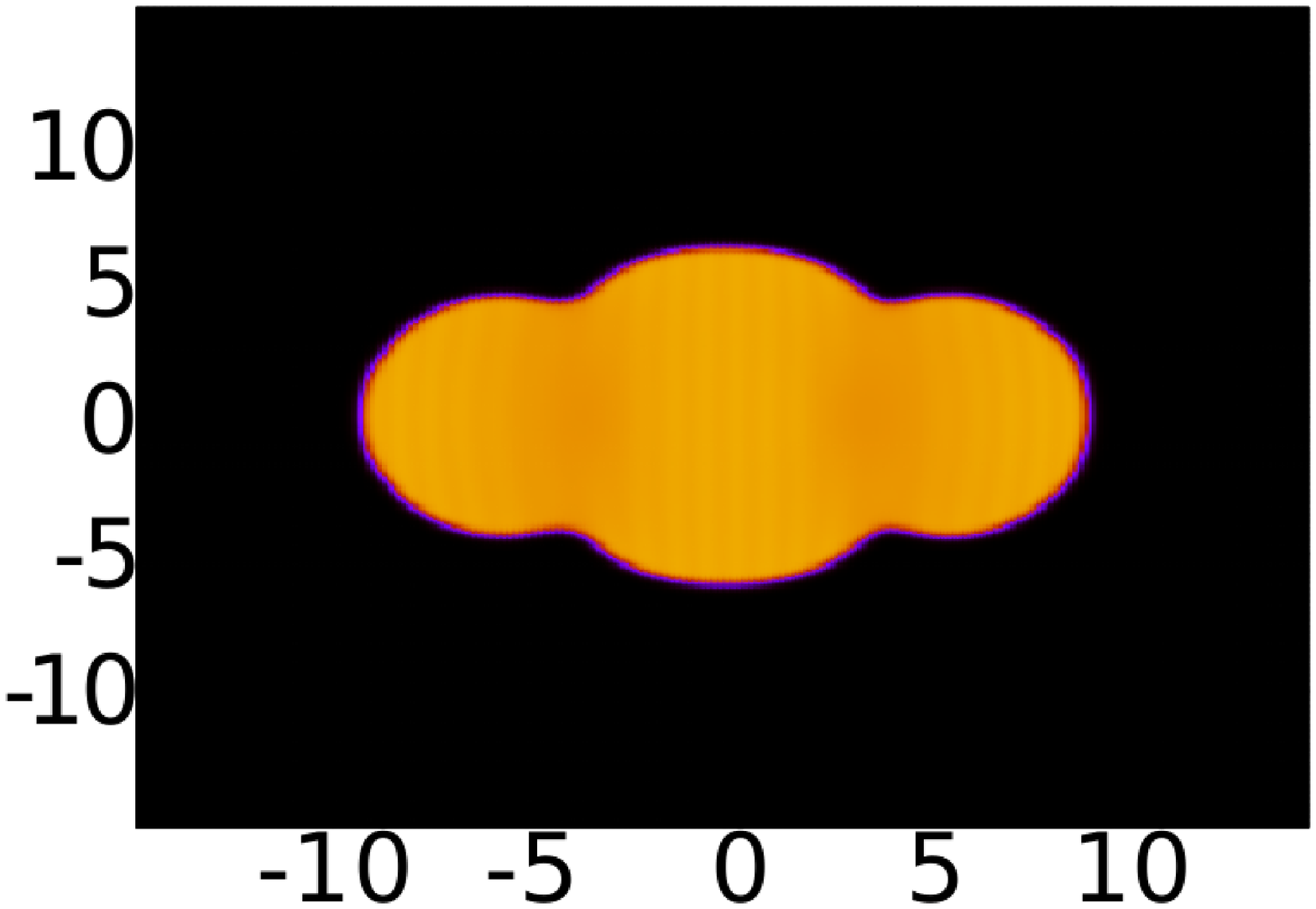}
              \includegraphics[height=1.75cm]{colorbar.eps}\\
       \includegraphics[width=2.5cm]{k1om1g50.eps}
       \includegraphics[width=2.5cm]{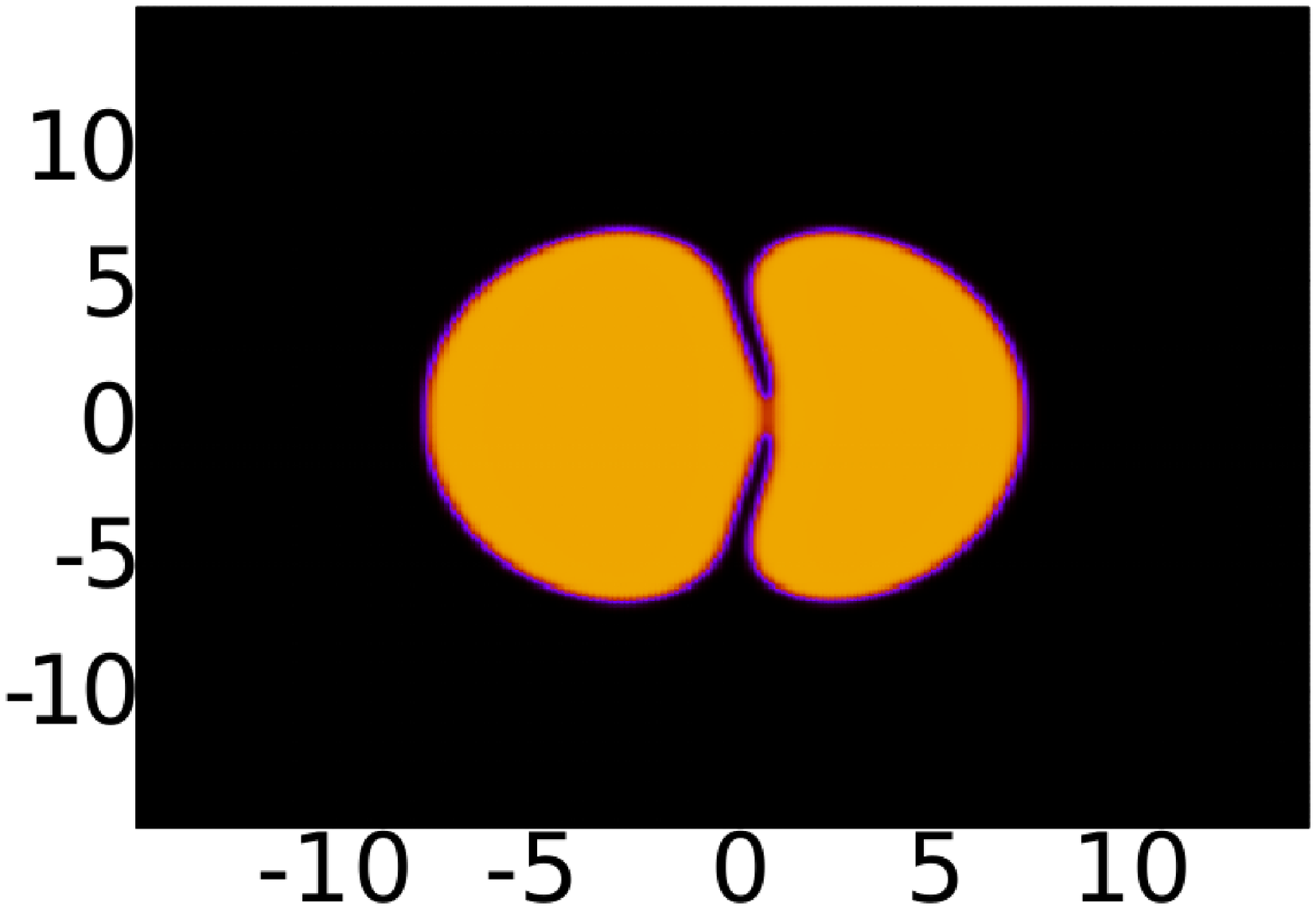}
       \includegraphics[width=2.5cm]{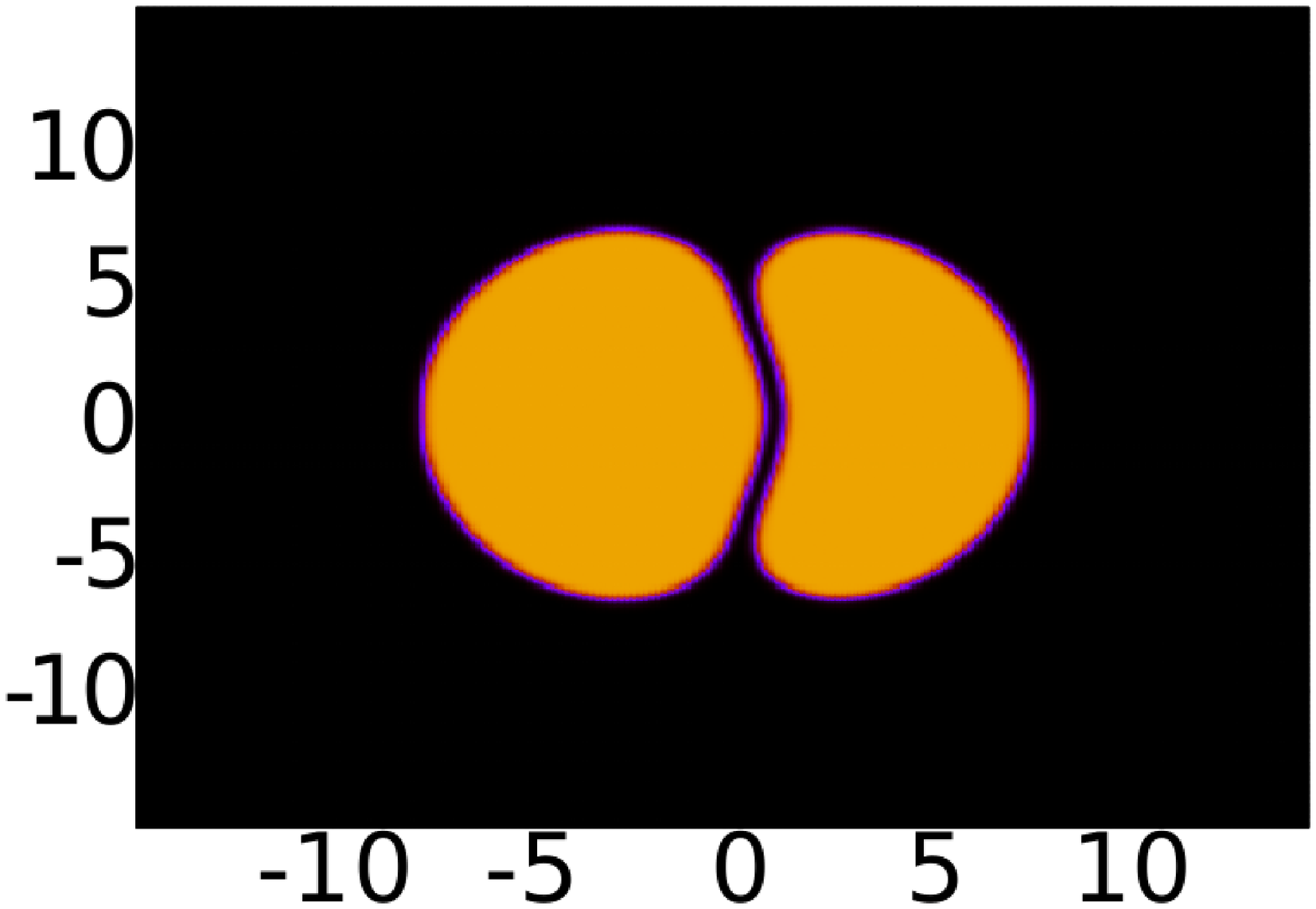} 
              \includegraphics[height=1.75cm]{colorbar.eps}\\
\caption{(color online) Density of the condensate in presence of Raman SOC interaction  for 100 number of atoms with different strength parameters.
{\bf 1st Row:} (Variation of the shape of the droplet with respect to interaction strength) $g = 20, 30, 50$ respectively. The other parameters are $k_0 = 1.0$ and $\Omega = 1.0$. 
{\bf 2nd Row:} (Variation with respect to SOC strength) $k_0 = 1.0, 3.0, 5.0$ respectively. The other parameters are $g = 50.0$ and $\Omega = 1.0$. 
{\bf 3rd Row:} (Variation with respect to Rabi frequency) $\Omega = 1.0, 10.0, 50.0$ respectively. The other parameters are $g = 50.0$ and $k_0 = 1.0$. }
    \label{ground_Raman}    
    \includegraphics[width=0.5\textwidth]{exc_ram_qx.eps}
     \includegraphics[width=0.5\textwidth]{exc_ram_qy.eps}   
\caption{Energy spectra two components of BEC in presence of Raman SOC. {\bf (a)} variation with respect to the two particle interaction strength. {\bf (b)} variation with strength of SOC {\bf (c)} variation with respect to Rabi frequency.}
\label{Raman_exc}
  \end{center}
\end{figure}

\small
\begin{eqnarray}
     i\frac{\partial\psi_1 }{\partial t}=\left[-\frac{\nabla ^2}{2} +  ik_0 \partial_x+ \frac{\delta}{2}+  V_{INT}   \right ]\psi_1+ \frac{\Omega}{2}\psi_2 + L_{HY}\psi_1\nonumber\\%\frac{g^2}{4\pi}(|\psi_1|^2+|\psi_2|^2) ln (|\psi_1|^2+|\psi_2|^2)\\
          i\frac{\partial\psi_2 }{\partial t}=\left[-\frac{\nabla ^2}{2}  - ik_0 \partial_x- \frac{\delta}{2}-V_{INT} \right ]\psi_2+ \frac{\Omega}{2}\psi_1 + L_{HY}\psi_2
 \end{eqnarray}
 \normalsize
where, $V_{INT} = g(|\psi_1|^2-|\psi_2|^2)$ is interaction part of the equations, $\delta$ is the detuning parameter (we have choose $\delta=0$ in our study), and $L_{HY}=\frac{g^2}{4\pi} \rho \; ln\left (\rho \right )$ is the quantum fluctuation part of the equations,  $\Omega$ is the Rabi frequency and $k_0$ is the strength of SOC. In the ultracold atomic system one can very all parameters in a controlled manner. Solving the above mentioned equation we get the density of the condensate, shown in the FIG \ref{ground_Raman}.
%figure 4

%After getting the ground state wave function we have applied Bogoliubov therory of excitation over the ground state. We have chosen the excited stete wave function as a perturbation over ground state like $\psi_j^{\mbox{exc}} = \psi_j+\delta \psi_j$. The perturbation part of the wave function can be written as $\delta \psi_j = U_j exp(iq.r -i\omega t) -V_j^* exp (-iq.r +i\omega^* t)$. If we put these in our GP equations we will find the equations for $\{U_j, V_j\}$
The excitation spectra can be obtained by diagonalizing the matrix (similarly as earlier)
%\begin{widetext}
\small 
  \begin{eqnarray}
\left(
    \begin{tabular}{c c c c}
      $H_{1}-k_0q_x$ & $\frac{\Omega}{2}+A {\psi}_1 {\psi}_2^*$ & $B {\psi}_1^2$&$ A{\psi}_1 {\psi}_2$\\
       $\frac{\Omega}{2}+A {\psi}_1^* {\psi}_2$ & $H_2+k_0q_x$ & $A {\psi}_1 {\psi}_2$ &  $B {\psi}_2^2$\\
       $-B{\psi}_1^{*2}$ & $-A{\psi}_1^* {\psi}_2^*$ &$-H_1-k_0q_x$ & $-(\frac{\Omega}{2}+A {\psi}_1^* {\psi}_2)$\\
       $-A {\psi}_1^* {\psi}_2^*$ & $-B {\psi}_2^{*2}$ & $-(\frac{\Omega}{2}+A {\psi}_1 {\psi}_2^*)$ & $-H_2+k_0q_x$\\
    \end{tabular}
\right)
\end{eqnarray}

\normalsize

The excitation spectrum are shown in Fig. \ref{Raman_exc} for various strength of SOC and Rabi frequencies.

\section{Results and discussions}

 {\Large a) } {\bf Circular film in absence of SOC: }
We have solved the GP equation in 2D BEC for the system of SOC atoms in the liquid phase. % Unlike the nonuniform confined system, we do not have any confinement potential, here the condensate is confined due to the attractive interactions between atoms.
 In this study, we have considered the repulsive interaction between the atoms of same species and attractive interaction between the atoms in different species. The quantum fluctuation is responsible for preventing the collapse of the condensate due to attractive interaction.
We get the circular liquid film for the system without SOC. It is clear from our calculation (left panel of FIG. \ref{2D_LHY}), if we increase the number of particles the size of the film increase keeping the density fixed. 
The collective excitation for such system is phonon-like mode for weak two body short range interaction and if we increase the strength of the interaction the mode becomes roton like, the position of the roton minimum shift to the larger momentum as we increase the strength of interaction, see the right panel of FIG. \ref{2D_LHY}. If we increase the strength of interaction  we will have two roton minia instead of one.

 {\Large b)} {\bf The film of atoms with Rashba and Dresselhaus SOC:} The film is nearly circular, not  precisely circular due to the Rashba SOC as shown in the  FIG. \ref{density_Rashba}. The density inside the film is constant over the entire region. There is very small asymmetry in collective excitation along X and Y-direction in momentum space. We have plotted the energy spectra for different Rashba SOC strength for different atom-atom interaction strengths in FIG 4(a). The variation of energy spectra with Rashba SOC has been shown in FIG 4(b) and 4(c) for $g=3$ and $g=15$ respectively. The nature of the excitation is almost same that of the system without SOC.
% ..............
  Phonon mode disappears in the presence of both the Rashba and   Dresselhaus SOC (see FIG. \ref{exc_RDSOC}) and we have roton mode of excitation even at small $g$.

 {\Large c)} {\bf The film of atoms with Raman SOC: }
The ground state of the BEC of atoms with Raman SOC separate into two centered (see FIG. \ref{ground_Raman}) due to the properties of single particle quantum state. The effective interaction gives a negative contribution to the total energy, whereas the kinetic energy (comes from the surface of the droplet, the surface tension) gives a positive contribution. The kinetic energy tries to evaporate the droplet whereas the interaction tries to bound the particles to form the droplet. If we increase the anisotropy by changing the interaction parameters, there is a great possibility of evaporation for anisotropic droplet due to the large surface area with high gradient. We have considered large number of particles and kept the parameters within the safe region of stable droplets. We have used any part of the condensate to get the collective excitation as the density and phase of the ground state is identical in the two regions. The excitation spectra are shown in FIG. \ref{Raman_exc}. The energy spectra are symmetric along Y-direction and asymmetric along X-direction of momentum space. 
\vskip -0.12cm
The energy spectrum has two roton minima for some parameters.
One important thing we have observed that the excitation becomes zero at a particular value of $k_0$ ($k_0\simeq 4.5$) and it becomes negative if we further increase the value of $k_0$. So the BEC collapse for large SOC. The long wavelength excitation and position of roton minima do not depend on the variation of $k_0$ but the roton-energy depends strongly on $k_0$.
\vskip -0.12cm
 Variation of excitation with the Rabi frequency, $\Omega$ for fixed $g$ ($g=50$) and SOC ($k_0=1$) is shown in the right part of FIG. \ref{Raman_exc}. Long wave lenght mode of the excitation changes abruptly with the Rabi frequency. For low value of $\Omega$ we have maximum of spectra at $q \rightarrow 0$, if we increase the value of $\Omega$ the magnitude of curvature reduces and becomes zero at $\Omega = 40$, after that we have positive curvature and minimum for higher value of Rabi frequency.

\subsection{Conclusions}

We have studied here the bulk properties corresponding to the interior of large droplets (in the form of film) for two different spin-orbit coupling systems. For simplicity we take the mixture of two BEC of atoms of same isotope of an element with different internal degrees of freedom.
    The droplet forms due to the particle-particle interaction and SOC makes the droplet asymmetric by deforming it. At small SOC the qualitative nature of the excitation remains identical with that of the without SOC. If we increase the strength of the SOC the droplet becomes unstable (which is obvious from the collective excitation) and evaporates due to the anisotropy.
  We have seen that the excitation contains phonon-mode for very small interacting system and otherwise, it contains roton-minima, whereas for gas phase we have both phonon and roton modes\cite{Stringari2012}. Some spectra contain  single roton minimum  and some contain double roton minima depending upon the strength of the interaction. In presence of Rashba-Dresselhaus SOC, we lost the phonon-mode of excitation even at small interaction. Roton energy strongly depends on $k_0$, the strength of Raman SOC, whereas the long-wavelength mode of excitation strongly depends on the Rabi frequency. We have very slow qualitative change in the nature of excitations inside the droplet due to the addition of SOC, we hope there will be abrupt qualitative changes of surface modes of excitation due to the change of shape of the droplet in presence of SOC, which is remained to study.

\section{acknowledgement}

DM thanks Subhasis Sinha (Department of Physical Sciences, IISER Kolkata) for the fruitful discussions.

%Bibliography .......

\end{document}